\documentclass[a4paper,notitlepage,12pt]{article}

\begin{document}
\title{Global transposable characteristics in the yeast complete DNA sequence}
\author{ Zuo-Bing Wu\footnotemark[1]\\
State Key Laboratory of Nonlinear Mechanics, \\
 Institute of Mechanics,\\
  Chinese Academy
of Sciences, Beijing 100190, China}
 \maketitle

\footnotetext[1]{Correspondence to: wuzb@lnm.imech.ac.cn}

\begin{abstract}
Global transposable characteristics in the complete DNA sequence
of the Saccharomyces cevevisiae yeast is determined by using the
metric representation and recurrence plot methods. In the form of
the correlation distance of nucleotide strings, 16 chromosome
sequences of the yeast, which are divided into 5 groups, display 4
kinds of the fundamental transposable characteristics: a short
period increasing, a long quasi-period increasing, a long major
value and hardly relevant.

\end{abstract}

\newpage
\section{Introduction}

The recent complete DNA sequences of many organisms are available
to systematically search of genome structure. For the large amount
of DNA sequences, developing methods for extracting meaningful
information is a major challenge for bioinformatics. To understand
the one-dimensional symbolic sequences composed of the four
letters `A', `C', `G' and `T' (or `U'), some statistical and
geometrical methods were
developed\cite{DEKM,Hao,QC,Li,RRP,SG,MAL,Li2,PO,GGS,AV}. In
special, chaos game representation (CGR)\cite{Jeffrey}, which
generates a two-dimensional square from a one-dimensional
sequence, provides a technique to visualize the composition of DNA
sequences. The characteristics of CGR images was described as
genomic signature, and classification of species in the whole
bacteria genome was analyzed by making an Euclidean metric between
two CGR images\cite{DGVFF}. Based on the genomic signature, the
distance between two DNA sequences depending on the length of
nucleotide strings was presented\cite{WHSK} and the horizontal
transfers in prokaryotes and eukaryotes were detected and
charaterized\cite{DFLGD,MBD}. Recently, a one-to-one metric
representation of the DNA sequences\cite{Wu1}, which was borrowed
from the symbolic dynamics, makes an ordering of subsequences in a
plane. Suppression of certain nucleotide strings in the DNA
sequences leads to a self-similarity of pattern seen in the metric
representation of DNA sequences. Self-similarity limits of genomic
signatures were determined as an optimal string length for
generating the genomic signatures\cite{Wu2}. Moreover, by using
the metric representation method, the recurrence plot technique of
DNA sequences was established and employed to analyze correlation
structure of nucleotide strings\cite{Wu3}.

As a eukaryotic organism, yeast is one of the premier industrial
microorganisms, because of its essential role in brewing, baking,
and fuel alcohol production. In addition, yeast has proven to be
an excellent model organism for the study of a variety of
biological problems involving the fields of genetics, molecular
biology, cell biology and other disciplines within the biomedical
and life sciences. In April 1996, the complete DNA sequence of the
yeast (Saccharomyces cevevisiae) genome, consisting of 16
chromosomes with 12 million basepairs, had been released to
provide a resource of genome information of a single organism.
However, only 43.3\% of all 6000 predicted genes in the
Saccharomyces cerevisiae yeast were functionally characterized
when the complete sequence of the yeast genome became
available\cite{MABFGHHKMOPZ}. Moreover, it was found that DNA
transposable elements have ability to move from place to place and
make many copies within the genome via the
transposition\cite{OLG,BEN}. Therefore, the yeast complete DNA
sequence remain a topic to be studied respect to its genome
architecture structure in the whole sequence.

In this paper, using the metric representation and recurrence plot
methods, we analyze global transposable characteristics in the
yeast complete DNA sequence, i.e., 16 chromosome sequences.

\section{Metric representation and recurrence plot methods}

For a given DNA sequence $s_1 s_2 \cdots s_i \cdots s_N$ ($s_i \in
\{A,C,G,T\}$), a plane metric representation is generated by
making the correspondence of symbol $s_i$ to number $\mu_i$ or
$\nu_i \in \{0,1\}$ and calculating values ($\alpha$, $\beta$) of
all subsequences $\Sigma_k=s_1 s_2 \cdots s_k$ ($1 \le k \le N$)
defined as follows
 \begin{equation}
 \begin{array}{l}
 \alpha  = 2\sum_{j=1}^k \mu_{k-j+1} 3^{-j} +3^{-k}= 2\sum_{i=1}^k \mu_i 3^{-(k-i+1)}
 +3^{-k},\\
 \beta = 2\sum_{j=1}^k \nu_{k-j+1} 3^{-j} +3^{-k}= 2\sum_{i=1}^k \nu_i 3^{-(k-i+1)} +3^{-k},
 \label{eq1}
 \end{array}
 \end{equation}
 where $\mu_i$ is 0 if $s_i \in \{A,C\}$ or 1 if $s_i \in \{G,T\}$
 and $\nu_i$ is 0 if $s_i \in \{A,T\}$ or 1 if $s_i \in \{C,G\}$.
 Thus, the one-dimensional symbolic sequence is partitioned into
 $N$ subsequences $\Sigma_k$ and mapped in the two-dimensional plane ($\alpha, \beta$).
 Subsequences with the
same ending $l$-nucleotide string, which are labeled by
$\Sigma^{l}$, correspond to points in the zone encoded by the
$l$-nucleotide string. Taking a subsequence $\Sigma_i \in
\Sigma^{l}$, we calculate
\begin{equation}
\Theta(\epsilon_{l}-|\Sigma_i-\Sigma_j|)= \Theta(\epsilon_{l}-
\sqrt{(\alpha_i-\alpha_j)^2+(\beta_i-\beta_j)^2}), \label{eq3}
\end{equation}
where $\Theta$ is the Heaviside function [$\Theta(x)=1$, if $x >
0$; $\Theta(x)=0$, if $x \leq 0$] and $\Sigma_j$ is a subsequence
($j \geq l$). When $\Theta(\epsilon_{l}-|\Sigma_i-\Sigma_j|)=1$,
i.e., $\Sigma_j \in \Sigma^{l}$, a point $(i,j)$ is plotted in a
plane. Thus, repeating the above process from the beginning of
one-dimensional
 symbolic sequence and shifting forward, we obtain a recurrence
 plot of the DNA sequence.

 For presenting correlation structure
 in the recurrence plot plane, a correlation intensity is defined
 at a given correlation distance $d$
\begin{equation}
\Xi(d) = \sum_{i=1}^{N-d}
\Theta(\epsilon_{l}-|\Sigma_i-\Sigma_{i+d}|). \label{eq4}
\end{equation}
The quantity displays the transference of $l$-nucleotide strings
in the DNA sequence. To further determine positions and lengths of
the transposable elements, we analyze the recurrent plot plane.
Since $\Sigma_i$ and $\Sigma_j$ $\in \Sigma^{l}$, the transposable
element has the length $l$ at least. From the recurrence plot
plane, we calculate the maximal value of $x$ to satisfy
\begin{equation}
\Theta(\epsilon_{l}-|\Sigma_{i+x}-\Sigma_{j+x}|)=1, \ \ \
x=0,1,2,\cdots \label{eq5}
\end{equation}
i.e., $\Sigma_{i+x}$ and $\Sigma_{j+x} \in \Sigma^l$.
 Thus, the transposable
element with the correction distance $d=j-i+1$ has the length
$L=l+x$. The transposable element is placed at the position
$(i-l+1, i+x)$ and $(j-l+1, j+x)$.

\section{Global transposable characteristics in the yeast complete DNA sequence}

The Saccharomyces cevevisiae yeast has 16 chromosomes, which are
denoted as YEAST I to XVI.
 Using the metric
representation and recurrence plot methods, we analyze correlation
structures of the 16 DNA sequences. According to the
characteristics of the correlation structures, we summarize the
results as follows:

(1) The correlation distance has a short period increasing. The
YEAST I, IX and XI have such characteristics. Let me take the
YEAST I as an example to analyze. Fig.1 displays the correlation
intensity at different correlation distance $d( \leq N-l)$ with
$l=15$. A local region is magnified in the figure. It is clearly
evident that there exist some equidistance parallel lines with a
basic correlation distance $d_b=135$. Using Eq. (4), we determine
positions and lengths of the transposable elements in Table I,
where their lengths are limited in $L \geq 100$. Many nucleotide
strings have correlation distance, which is the integral multiple
of $d_b$. They mainly distribute in two local regions of the DNA
sequence (25715-26845) and (204518-206554) or (11.2-11.7\%) and
(88.8-89.7\%) expressed as percentages. The YEAST IX and XI have
similar behaviors. The YEAST IX has the basic correlation distance
$d_b=18$. Many nucleotide strings ($L \geq 50$) with the integral
multiple of $d_b$ mainly distribute in a local region of the DNA
sequence (391337-393583) or (89.0-89.5\%) expressed as
percentages. The YEAST XI has the basic correlation distance
$d_b=189$. Many nucleotide strings ($L \geq 50$) with the integral
multiple of $d_b$ mainly distribute in a local region of the DNA
sequence (647101-647783) or (97.1-97.2\%) expressed as
percentages.

(2) The correlation distance has a long major value and a short
period increasing. The YEAST II, V, VII, VIII, X, XII, XIII, XIV,
XV and XVI have such characteristics. Let me take the YEAST II as
an example to analyze. Fig.2 displays the correlation intensity at
different correlation distance $d( \leq N-l)$ with $l=15$. The
maximal correlation intensity appears at the correlation distance
$d_m=38534$. A local region is magnified in the figure. It is
clearly evident that there exist some equidistance parallel lines
with a basic correlation distance $d_b=36$. In Table II, positions
and lengths ($L \geq 50$) of the transposable elements are given.
The maximal transposable elements mainly distribute in two local
regions of the DNA sequence (221249-224565, 259783-263097) or
(27.2-27.6\%, 31.9- 32.4\%) expressed as percentage. Near the
positions, there also exist some transposable elements with
approximate values for $d_b$. Moreover, many nucleotide strings
have correlation distance, which is the integral multiple of
$d_b$. They mainly distribute in a local region of the DNA
sequence (391337-393583) or (89.0-89.5\%) expressed as
percentages.
 In the other 9 DNA sequences, the YEAST V, X,
XII, XIII, XIV, XV and XVI have the same basic correlation
distance $d_b=36$ and similar behaviors with different major
correlation distance $d_m=$49099, 5584, 9137, 12167, 5566, 447110
and 45988, respectively. The YEAST VII and VIII have different
basic correlation distance $d_b=12$ and 135, and similar behaviors
with the major correlation distance $d_m=255548$ and 1998,
respectively.

(3) The correlation distance has a long quasi-period increasing.
The YEAST III has such characteristics. Fig. 3 displays the
coherence intensity at different correlation distance $d( \leq
N-l)$ with $l=15$. The correlation intensity has the maximal value
at the correlation distance $d_{m1}=185903$ and two vice-maximal
values at the correlation distance $d_{m2}=93625$ and
$d_{m3}=279528$. Since $d_{m2} \approx d_{m1}/2 \approx d_{m3}/3$,
the coherence distance has a quasi-period increasing. A local
region is magnified in the figure. These does not exist any clear
short period increasing of the correlation distance. Using Eq.
(4), we determine positions and lengths ($L \geq 50$) of the
transposable elements in Table III. The maximal and vice-maximal
transposable elements mainly distribute in local regions of the
DNA sequence (11499-13810, 197402-199713), (198171-199796,
291794-293316) and (12268-12932, 291794-292460) or (3.6-4.4\%,
62.6-63.6\%), (62.8-63.4\%, 92.5-93.0\%) and (3.9-4.1\%,
92.5-92.7\%) expressed as percentage.

(4) The correlation distance has a long major value and a long
quasi-period and two short period increasing. The YEAST IV has
such characteristics. Fig. 4 displays the coherence intensity at
different correlation distance $d( \leq N-l)$ with $l=15$. The
maximal coherence intensity appears at the correlation distance
$d_m=3885$. There also exist three vice-maximal values at the
correlation distance $d_{m1}=232800$, $d_{m2}=109349$ and
$d_{m3}=341221$, which forms a long quasi-period increasing of the
correlation distance, i.e., $d_{m2} \approx d_{m1}/2 \approx
d_{m3}/3$. A local region is magnified in the figure. It is
clearly evident that there exist two short period increasing with
$d_{b1}=84$ and $d_{b2}=192$ in the correlation distance. In Table
IV, positions and lengths ($L \geq 100$) of the transposable
elements are determined by using Eq. (4). All correlation distance
with the long major value and the long quasi-period and two short
period increasing are denoted. The transposable elements with
$d_m$, $d_{m1}$, $d_{m2}$, $d_{m3}$, $d_{b1}$ and $d_{b2}$ mainly
distribute in local regions of the DNA sequence (527570-538236),
(871858-876927, 981207-986276), (645646-651457, 878346-884257),
(646379-651032, 987600-992253), (1307733-1308591) and
(758135-759495) or (34.4-35.1\%), (56.9-57.2\%, 64.0-64.4\%),
(42.1-42.5\%, 57.3-57.7\%), (42.2-42.5\%, 64.4-64.8\%),
(85.36-85.41\%) and (49.5-49.6\%) expressed as percentages.

 (5) The DNA
sequence is hardly relevant. The YEAST VI has such
characteristics. Fig. 5 displays the coherence intensity at
different correlation distance $d( \leq N-l)$ with $l=15$. The
maximal coherence intensity appears at the correlation distance
$d_m=5627$. A local region is magnified in the figure. The
sequence has not a short period increasing of the coherence
distance. In Table V, positions and lengths ($L \geq 50$) of the
transposable elements are given. Only one nucleotide string with
the length 337 has the correlation distance $d_m$. The YEAST VI is
almost never relevant, so the YEAST VI approaches a random
sequence.

\section{Conclusion}
Global transposable characteristics in the yeast complete DNA
sequence is determined by using the metric representation and
recurrence plot methods. Positions and lengths of all transposable
nucleotide strings in the 16 chromosome DNA sequences of the yeast
are determined. In the form of the correlation distance of
nucleotide strings, the fundamental transposable characteristics
displays a short period increasing, a long quasi-period
increasing, a long major value and hardly relevant. The 16
chromosome sequences are divided into 5 groups, which have one or
several of the 4 kinds of the fundamental transposable
characteristics.

\textbf{Acknowledgments} We thank the IMECH and ICTS research
computing facilities for assisting us in the computation.

\newpage
{\small
 Table I. Transference of nucleotide strings with lengths
$L( \geq 100)$ for the YEAST I with 230209 bases.

\begin{tabular}{lllllll}
\hline
No. & String  & Position 1     & Position 2 & $L$ & $d$  & Note  \\
\hline
1   & $t^2a \cdots act$ &   11745-11969 &   24177-24401 &   225 &   12432   &\\
2   & $ctg \cdots a^2t$ &   12258-12396 &   24711-24849 &   139 &   12453   &\\
3   & $g^2a \cdots g^2a$ &  12988-13171 &   25153-25336 &   184 &   12165   &\\
4   & $c^2g \cdots cgt$ &   25715-25851 &   26255-26391 &   137 &   540     & $4d_b$\\
5   & $at^2 \cdots ac^2$  &  25739-25851 & 26414-26526 &   113 &   675  & $5d_b$\\
6   & $gta \cdots ac^2$  &  25751-25851 & 26561-26661 & 101 &   810  & $6d_b$\\
7   & $gta \cdots ac^2$  & 25751-25851 &   26696-26796 & 101 &   945  & $7d_b$\\
8   & $t^2g \cdots g^2t$ &  25853-25968 & 26393-26508 &   116 &   540  & $4d_b$\\
9   & $atg \cdots gtg$ & 25925-26035 & 26060-26170 & 111 &   135  & $d_b$ \\
10  & $atg \cdots agt$ &  25925-26058 & 26195-26328 & 134 &   270  & $2d_b$ \\
11  & $agt \cdots gtg$ & 26050-26170 &   26185-26305 & 121 & 135  & $d_b$ \\
12  & $at^2 \cdots gac$ & 26279-26406 & 26414-26541 & 128 &   135  & $d_b$ \\
13  & $gta \cdots gac$ & 26291-26406 & 26561-26676 & 116 & 270  & $2d_b$ \\
14  & $gta \cdots gac$ & 26291-26406 & 26696-26811 & 116 & 405  & $3d_b$ \\
15  & $gta \cdots gtg$ & 26426-26710 & 26561-26845 &   285 & 135  & $d_b$ \\
16  & $tga \cdots aca$ & 160239-160575 & 165827-166163 & 337 &   5588 &\\
17  & $cac \cdots tac$ & 204518-204802 & 204653-204937 &   285 & 135  & $d_b$ \\
18  & $g^2t \cdots tac$ & 204567-204667 & 205512-205612 &   101 & 945   & $7d_b$\\
19  & $g^2t \cdots tac$ & 204702-204802 & 205512-205612 & 101 & 810   & $6d_b$\\
20  & $g^2t \cdots a^2t$ & 204837-204949 & 205512-205624 & 113 & 675   & $5d_b$\\
21  & $ac^2 \cdots c^2a$ & 204855-204969   & 205395-205509 & 115 & 540   & $4d_b$\\
22  & $ctc \cdots cat$ & 205042-205168   & 205312-205438   & 127 & 270   & $2d_b$\\
23  & $cac \cdots act$ & 205058-205178 & 205193-205313 & 121 &   135   & $d_b$\\
24  & $cac \cdots cat$ & 205193-205303 & 205328-205438 &   111 & 135   & $d_b$\\
25  & $atg \cdots t^2c$ & 205758-205879 & 206433-206554 &   122 & 675  & $5d_b$\\
 \hline
\end{tabular}
}

\newpage
{\small
 Table II. Transference of nucleotide strings with lengths
$L( \geq 50)$ for the YEAST II with 813142 bases. Due to the
limited spacing, 22 nucleotide strings in the total number 57 are
not presented.

\begin{tabular}{lllllll}
\hline
No. & String &  Position 1     & Position 2 & $L$ & $d$   & Note\\
\hline
1   & $tag \cdots agt$ &  1584-1650   &   1692-1758   &   67  &   108 & $3d_b$\\
2   & $tag \cdots gct$ & 1620-1674   &   2016-2070   &   55  &   396 & $11d_b$\\
3   & $tag \cdots gct$ &  1620-1674 & 2268-2322   &   55  &   648 & $18d_b$\\
4   & $tgc \cdots tg^2$ & 1815-1870 &   1851-1906 & 56  &   36 & $d_b$\\
5   & $gct \cdots gta$ &  1860-1936   & 1932-2008   & 77  &   72 & $2d_b$\\
6   & $tgc \cdots gta$ & 1887-1936   & 2355-2404   &   50  & 468 & $13d_b$\\
7   & $tgc \cdots gtg$ & 1959-2018 & 2355-2414   &   60  & 396 & $11d_b$\\
8   & $gca \cdots agt$ & 2012-2082 & 2264-2334 & 71 &   252 & $7d_b$\\
9   & $tg^2 \cdots agt$ &  2022-2190   & 2094-2262   &   169 & 72 & $2d_b$\\
10  & $cag \cdots gta$ & 2037-2104 & 2325-2392   &   68 & 288 & $8d_b$\\
11  & $tg^2 \cdots agt$ & 2094-2154 & 2274-2334   & 61  & 180 & $5d_b$\\
12  & $cag \cdots gta$ & 2109-2176   & 2325-2392 &   68 & 216 & $6d_b$\\
13  & $tg^2 \cdots agt$ & 2166-2226 & 2274-2334   & 61 & 108 & $3d_b$\\
14  & $cag \cdots gta$ & 2181-2248   & 2325-2392   &   68 & 144 & $4d_b$\\
33  & $a^2t \cdots c^2t$ & 221182-221231 & 259718-259767 & 50  & 38536 & $\approx d_m$\\
35  & $gta \cdots tct$ & 221249-221308 & 259783-259842   & 60  & 38534 & $d_m$\\
37  & $gat \cdots tca$ & 222669-222827 & 261203-261361 & 159 & 38534 & $d_m$\\
38  & $ta^2 \cdots tgc$ & 222829-223096 & 261363-261630 &   268 & 38534 & $d_m$\\
39  & $cat \cdots gct$ & 223098-223609 & 261632-262143 &   512 & 38534 & $d_m$\\
40  & $a^2c \cdots g^2a$ & 223611-223966 & 262145-262500 &   356 & 38534 & $d_m$\\
41  & $tc^2 \cdots c^2g$ & 223968-224563 & 262502-263097 &   596 & 38534 & $d_m$\\
42  & $ga^2 \cdots tac$ & 224736-225308 & 263273-263845 &   573 & 38537 & $\approx d_m$\\
43  & $ata \cdots aca$ & 225310-225493 & 263847-264030 &   184 & 38537 & $\approx d_m$\\
44  & $ac^2 \cdots a^2c$ & 225841-226231 & 264378-264768 &   391 & 38537 & $\approx d_m$\\
45  & $acg \cdots t^3$ & 226233-226797 & 264770-265334 &   565 & 38537 & $\approx d_m$\\
\hline
\end{tabular}
}

\newpage
{\small

Table III. Transference of nucleotide strings with lengths $k(
\geq 50)$ for the YEAST III with 315341 bases.

\begin{tabular}{lllllll}
\hline
No. & String &  Position 1     & Position 2 & $L$ & $d$    & Note\\
\hline
1   & $c^3 \cdots ac^2$ &  90-141  &   233-284 &   52  &   143 &\\
2   & $a^2t \cdots ca^2$ &   1190-1253 & 4083-4146   &   64  &   2893 &\\
3   & $t^2g \cdots gta$ & 11499-11764 & 197402-197667 &   266 &   185903 & $d_{m1}$ \\
4   & $tag \cdots cta$ & 11766-11908 & 197669-197811   &   143 &   185903 & $d_{m1}$ \\
5   & $ta^2 \cdots gac$ & 11910-12185 & 197813-198088   &   276 & 185903 & $d_{m1}$ \\
6   & $at^2 \cdots gtg$ & 12187-13810 & 198090-199713   &   1624    & 185903 & $d_{m1}$ \\
7   & $tac \cdots tat$ & 12268-12340 & 291794-291866   &   73  &   279526 & $\approx d_{m3}$ \\
8   & $ata \cdots at^2$ & 12325-12932 & 291853-292460   &   608 &   279528 & $d_{m3}$  \\
9   & $a^2t \cdots gtg$ & 13691-13810 & 293114-293233   &   120 &   279423  & \\
10  & $cg^2 \cdots ca^2$ & 13812-14006 & 199715-199909   &   195 &   185903 & $d_{m1}$\\
11  & $cg^2 \cdots t^2c$ & 13812-13893 & 293235-293316   &   82  &   279423 &\\
12  & $tgt \cdots a^2c$ & 83677-83802 & 84419-84544 &   126 &   742 &\\
13  & $tgt \cdots a^2c$ & 83677-83802 & 90049-90174 & 126 &   6372 &\\
14  & $cta \cdots a^2c$ & 83724-83802 & 83951-84029 & 79  &   227 &\\
15  & $ca^2 \cdots tg^2$ & 83804-83902 & 84031-84129 &   99 &   227 &\\
16  & $ca^2 \cdots tg^2$ & 83804-83902 & 84546-84644 & 99  & 742 &\\
17  & $ca^2 \cdots tg^2$ & 83804-83902 & 90176-90274 & 99 & 6372 &\\
18  & $cta \cdots tct$ & 83951-84230 & 84466-84745 &   280 & 515 &\\
19  & $cta \cdots tat$ & 83951-84189 & 90096-90334 & 239 & 6145 &\\
20  & $tgt \cdots tat$ & 84419-84704 & 90049-90334 &   286 & 5630 &\\
21  & $cac \cdots t^3$ & 123942-124058 & 142661-142777 & 117 &   18719 &\\
22  & $tac \cdots tat$ & 198171-198243 & 291794-291866   &   73 & 93623  & $\approx d_{m2}$\\
23  & $ata \cdots at^2$ & 198228-198835 & 291853-292460 &   608 & 93625  & $d_{m2}$\\
24  & $a^2t \cdots t^2c$ & 199594-199796 & 293114-293316   &   203 & 93520  & $\approx d_{m2}$\\
25  & $g^2a \cdots g^3$ & 267365-267574 & 267692-267901   &   210 &   327 &\\
\hline
\end{tabular}
}

\newpage
{\small Table IV.  Transference of nucleotide strings with lengths
$L( \geq 100)$ for the YEAST IV with 1531977 bases. Due to the
limited spacing, 112 nucleotide strings without any notes in the
total number 176 are not presented.

\begin{tabular}{lllllll}
\hline
No. & String &  Position 1     & Position 2 & $L$ & $d$  & Note  \\
\hline
28  & $gct \cdots a^3$ &  527570-527835   &   531455-531720   &   266 &   3885    & $d_m$\\
29  & $gct \cdots a^2t$ &  527570-527781   &   535340-535551   &   212 &   7770    &\\
30  & $gt^2 \cdots c^2a$ &  527891-528066   &   531776-531951   &   176 &   3885    & $d_m$\\
31  & $gt^2 \cdots c^2a$ &  527891-528066   &   535661-535836   &   176 &   7770   & \\
32  & $ag^2 \cdots a^2t$ &  528094-531666   &   531979-535551   &   3573    &   3885   & $d_m$ \\
35  & $ct^2 \cdots t^2g$ &  531668-532364   &   535553-536249   &   697 &   3885   & $d_m$ \\
36  & $gca \cdots c^2a$ &  532366-534351   &   536251-538236   &   1986    &   3885  & $d_m$  \\
40  & $ata \cdots agt$ &  645546-646118   &   878346-878918   &   573 &   232800  & $d_{m2}$\\
41  & $ag^2 \cdots aca$ &  645635-645873   &   992440-992678   &   239 &   346805  &\\
42  & $agc \cdots atc$ &  646120-646336   &   878920-879136   &   217 &   232800  & $d_{m2}$\\
43  & $ct^2 \cdots gta$ &  646338-646726   &   879138-879526   &   389 &   232800  & $d_{m2}$\\
44  & $t^3 \cdots atg$ &  646379-646564   &   987600-987785   &   186 &   341221  & $d_{m3}$\\
45  & $t^2c \cdots cat$ &  646787-647179   &   879587-879979   &   393 &   232800  & $d_{m2}$\\
46  & $tat \cdots t^2c$ &  646964-647288   &   988185-988509   &   325 &   341221  & $d_{m3}$\\
47  & $t^3 \cdots ta^2$ &  647310-647470   &   880110-880270   &   161 &   232800  & $d_{m2}$\\
48  & $ta^2 \cdots tga$ &  647468-647693   &   988689-988914   &   226 &   341221  & $d_{m3}$\\
49  & $gt^2 \cdots cgt$ &  647472-648042   &   880272-880842   &   571 &   232800  & $d_{m2}$\\
50  & $gag \cdots aga$ &  647695-649519   &   988916-990740   &   1825    &   341221  & $d_{m3}$\\
51  & $gat \cdots cg^2$ &  648165-648265   &   880965-881065   &   101 &   232800  & $d_{m2}$\\
52  & $tc^2 \cdots atg$ &  648486-649715   &   881286-882515   &   1230    &   232800  & $d_{m2}$\\
53  & $cgt \cdots atg$ &  649521-649715   &   990742-990936   &   195 &   341221 & $d_{m3}$ \\
54  & $ctg \cdots t^3$ &  649853-650096   &   991074-991317   &   244 &   341221  & $d_{m3}$\\
55  & $tag \cdots ctc$ &  649946-650073   &   882746-882873   &   128 &   232800 & $d_{m2}$ \\
56  & $gat \cdots aca$ &  650075-651457   &   882875-884257   &   1383    &   232800  & $d_{m2}$\\
57  & $ctg \cdots ct^2$ &  650098-651032   &   991319-992253   &   935 &   341221 & $d_{m3}$ \\
 \hline
\end{tabular}
}

{\small
\begin{tabular}{lllllll}
\hline
60  & $ag^2 \cdots aca$ &  651219-651457   &   992440-992678   &   239 &   341221  & $d_{m3}$\\
63  & $ac^2 \cdots c^2a$ &  757478-757581   &   757670-757773   &   104 &   192 & $d_{b2}$\\
68  & $cta \cdots act$ &  758135-758259   &   758519-758643   &   125 &   384 & $2d_{b2}$\\
69  & $cta \cdots act$ &  758135-758259   &   758711-758835   &   125 &   576 & $3d_{b2}$\\
70  & $cta \cdots act$ &  758135-758259   &   759479-759603   &   125 &   1344  &  $7d_{b2}$\\
71  & $gca \cdots g^2a$ &  758219-758343   &   758411-758535   &   125 &   192 & $d_{b2}$\\
72  & $gca \cdots a^2t$ &  758219-758347   &   759179-759307   &   129 &   960 & $5d_{b2}$\\
73  & $gca \cdots g^2a$ &  758219-758343   &   759371-759495   &   125 &   1152  &  $6d_{b2}$\\
74  & $ctg \cdots act$ &  758349-758451   &   758925-759027   &   103 &   576 & $3d_{b2}$\\
75  & $ctg \cdots g^2a$ &  758349-758535   &   759117-759303   &   187 &   768 & $4d_{b2}$\\
76  & $ctg \cdots cat$ &  758349-758683   &   759309-759643   &   335 &   960 & $5d_{b2}$\\
77  & $cta \cdots cat$ &  758495-758683   &   758687-758875   &   189 &   192 & $d_{b2}$\\
78  & $cta \cdots aga$ &  758687-758901   &   759455-759669   &   215 &   768 & $4d_{b2}$\\
79  & $gca \cdots act$ &  758795-759027   &   758987-759219   &   233 &   192 & $d_{b2}$\\
80  & $cta \cdots act$ &  758903-759027   &   759287-759411   &   125 &   384 & $2d_{b2}$\\
81  & $gca \cdots aga$ &  758987-759093   &   759563-759669   &   107 &   576 & $3d_{b2}$\\
82  & $cta \cdots g^2a$ &  759095-759303   &   759287-759495   &   209 &   192 & $d_{b2}$\\
99  & $tgt \cdots a^3$ &  871858-872030   &   981207-981379   &   173 &   109349 & $d_{m1}$ \\
103 & $tgc \cdots ca^2$ &  872202-872307   &   981551-981656   &   106 &   109349   & $d_{m1}$ \\
104 & $a^2g \cdots cag$ &  872309-872592   &   981658-981941   &   284 &   109349   & $d_{m1}$ \\
105 & $at^2 \cdots tat$ & 872737-872871   &   982086-982220   &   135 &   109349   & $d_{m1}$ \\
106 & $ag^2 \cdots ca^2$ &  873022-873286   &   982371-982635   &   265 &   109349   & $d_{m1}$ \\
107 & $tga \cdots cat$ &  873378-874087   &   982727-983436   &   710 &   109349   & $d_{m1}$ \\
108 & $tca \cdots g^2t$ &  874089-874236   &   983438-983585   &   148 &   109349   & $d_{m1}$ \\
109 & $tac \cdots tgc$ &  874238-874593   &   983587-983942   &   356 &   109349   & $d_{m1}$ \\
110 & $aga \cdots atc$ &  874595-874764   &   983944-984113   &   170 &   109349   & $d_{m1}$ \\
111 & $t^3 \cdots ca^2$ &  874853-875247   &   984202-984596   &   395 &   109349   & $d_{m1}$ \\
112 & $gat \cdots a^2c$ &  875249-875604   &   984598-984953   &   356 &   109349   & $d_{m1}$ \\
 \hline
\end{tabular}

\begin{tabular}{lllllll}
\hline
113 & $ca^2 \cdots tga$ &  875637-876474   &   984986-985823   &   838 &   109349   & $d_{m1}$ \\
114 & $agc \cdots tga$ &  876491-876927   &   985840-986276   &   437 &   109349   & $d_{m1}$ \\
117 & $g^2a \cdots aga$ &  877085-877385   &   986434-986734   &   301 &   109349  &$d_{m1}$\\
118 & $ga^2 \cdots a^3$ &  877387-877657   &   986736-987006   &   271 &   109349  &$d_{m1}$\\
169 & $cgt \cdots ac^2$ &  1307733-1307835 &   1308321-1308423 &   103 &   588 & $7d_{b1}$ \\
170 & $ac^2 \cdots g^2c$ &  1307749-1307874 &   1308505-1308630 &   126 &   756 & $9d_{b1}$  \\
171 & $gt^2 \cdots atc$ &  1307753-1307871 &   1307921-1308039 &   119 &   168 & $2d_{b1}$  \\
172 & $gt^2 \cdots atc$ &  1307921-1308039 &   1308509-1308627 &   119 &   588 & $7d_{b1}$  \\
173 & $gt^2 \cdots atc$ &  1308089-1308249 &   1308341-1308501 &   161 &   252 & $3d_{b1}$  \\
174 & $ac^2 \cdots atc$ &  1308169-1308333 &   1308253-1308417 &   165 &   84  & $d_{b1}$  \\
175 & $ac^2 \cdots ac^2$ &  1308337-1308507 &   1308421-1308591 &   171 &   84 & $d_{b1}$   \\
\hline
\end{tabular}
}

\newpage
{\small
 Table V. Transference of nucleotide strings with
lengths $L( \geq 50)$ for the YEAST VI with 270148 bases.

\begin{tabular}{lllllll}
\hline
No. & String & Position 1     & Position 2 & $L$ & $d$   & Note \\
\hline

1   & $tat \cdots aca$ &  137905-138238   &   143532-143865   &   334 &   5627 & $d_m$\\
2   & $ga^2 \cdots t^3$ & 178016-178086   &   178157-178227   &   71  &   141 &\\
3   & $ca^2 \cdots gtc$ & 178088-178157   &   178229-178298   &   70  &   141 &\\
4   & $tgt \cdots gtg$ & 210332-210391   &   210334-210393   &   60  &   2 &\\
\hline
\end{tabular}
}

\newpage
\textbf{Figure caption}

Fig.~1. A plot of correlation intensity $\Xi(d)$ versus
correlation distance $d$ for the YEAST I.

Fig.~2. A plot of correlation intensity $\Xi(d)$ versus
correlation distance $d$ for the YEAST II.

Fig.~3. A plot of correlation intensity $\Xi(d)$ versus
correlation distance $d$ for the YEAST III.

Fig.~4. A plot of correlation intensity $\Xi(d)$ versus
correlation distance $d$ for the YEAST IV.

Fig.~5. A plot of correlation intensity $\Xi(d)$ versus
correlation distance $d$ for the YEAST VI.

\newpage
\begin{thebibliography}{}


\bibitem{DEKM} R. Durbin, S.R. Eddy, A. Krogh, and G. Mitchison.
{\it Probabilistic models of proteins and nucleic acids} (1999),
Cambridgr University Press.

\bibitem{Hao} B.-L. Hao, H. C. Lee, and S.-Y. Zhang.
Fractals related to long DNA sequences and bacterial complete
genomes. {\it Chaos, Solitons and Fractals} {\bf 11} (2000) 825.

\bibitem{QC} D. Qi, A. J. Cuticchia,
Compositional symmetries in complete genomes. {\it Bioinformatics}
{\bf 17} (2001) 557.

\bibitem{Li} W. Li, P. Bernaola-Galv\'an, F. Haghighi, I. Grosse.
 Applications of recursive segmentation to the analysis of DNA sequences.
 {\it Computers \& Chemistry} {\bf 26} (2002) 491.

\bibitem{RRP} D. Robelin, H. Richard, B. Prum, SIC: a tool
to detect short inverted segments in a biological sequence, {\it
Nucleic Acids Research} {\bf 31} (2003) 3669.

\bibitem{SG} S. Garte, Fractal properties of the human genome.
{\it Journal of Theoretical Biology} {\bf 230} (2004) 251.

\bibitem{MAL} P. W. Messer, P. F. Arndt, M. L\"assig (2005),
Solvable sequence evolution models and genomic correlations, {\it
Phys. Rev. Lett.} {\bf 94} (2005) 138103.


\bibitem{Li2} W. Li, P. Miramontes, Large-scale oscillation of
structure-related DNA sequence features in human chromosome 21,
{\it Physical Review E} {\bf 74} (2006) 021912.

\bibitem{PO} A. Provata, Th. Oikonomou, Power law exponents
characterizing human DNA, {\it Physical Review E} {\bf 75} (2007)
056102.

\bibitem{GGS} H. Gonzalez-Diaz, Y. Gonzalez-Diaz, L. Santana, et al.
Proteomics, networks and connectivity indices, {\it Proteomics}
{\bf 8} (2008) 750.

\bibitem{AV} J. S. Almeida and S. Vinga,
Biological sequences as pictures - a generic two dimensional
solution for iterated maps, {\it BMC Binoinformatics} {\bf 10}
(2009) 100.

\bibitem{Jeffrey} H. J. Jeffrey, Chaos game representation of gene structure.
 {\it Nucleic Acids Res.} {\bf 18} (1990) 2163.

 \bibitem{DGVFF} P. J. Deschavanne, A. Giron, J. Vilain, G. Fagot, and B. Fertil,
 Genomic signature: characterization and classification of species assessed
 by chaos game representation of sequences.
 {\it Mol. Biol. Evol.} {\bf 16} (1999) 1391.

\bibitem{WHSK} Y. Wang, K. Hill, S. Singh and L. Kari, The
spectrum of genomic signatures: from dinucleotides to chaos game
representation. {\it Gene} {\bf 346} (2005) 173.

\bibitem{DFLGD} C. Dufraigne, B. Feitil, S. Lespinats, A. Giron
and P. Deschavanne, Detection and characterization of horizontal
transfers in prokaryotes using genomic signature. {\it Nucleic
Acids Res.} {\bf 33} (2005) e6.

\bibitem{MBD} L. V. Mallet, J. Becq and P. Deschavanne,
Whole genome evaluation of horizontal transfers in the pathogenic
fungus Aspergillus fumigatus. {\it BMC Genomics} {\bf 11} (2010)
171.

\bibitem{Wu1} Z.-B. Wu, Metric representation of DNA sequences.
{\it Electrophoresis} {\bf 21} (2000) 2321.

\bibitem{Wu2} Z.-B. Wu, Self-similarity limits of genomic signatures.
{\it Fractals} {\bf 11} (2003) 19.

\bibitem{Wu3} Z.-B. Wu, Recurrence plot analysis of DNA sequences.
{\it Phys. Lett. A} {\bf 232} (2004) 250.

\bibitem{MABFGHHKMOPZ} H. W. Mewes, K. Albermann, M. B\"ahr et
al. Overview of the yeast genome, {\it Nature} {\bf 387} (1997) 7.

\bibitem{OLG} H. Ochman, J. G. Lawrence and E. A. Groisman,
Lateral gene transfer and the nature of bacterial innovation, {\it
Nature} {\bf 405} (2000) 299.

\bibitem{BEN} J. L. Bennetzen, Transposable element contributions
to plane gene and genome evolution, {\it Plant Mol. Biol.} {\bf
42} (2000) 251.

\end{thebibliography}
\end{document}